\documentclass[a4 paper, 12 pt] {article}

\def\be{\begin{equation}}
\def\ee{\end{equation}}
\def\bdi{\begin{displaymath}}
\def\edi{\end{displaymath}}
\def\br{\begin{eqnarray}}
\def\er{\end{eqnarray}}

\def\RR{{\rm I\kern-.1567em R}}                              
 \def\CC{{\rm C\kern-4.7pt                                    
 \vrule height 7.7pt width 0.4pt depth -0.5pt \phantom {.}}}
 \def\ZZ{{\sf Z\kern-4.5pt Z}}                                

\begin{document}

\begin{titlepage}
\vspace*{-2 cm}
\noindent

\vskip 3cm
\begin{center}
{\Large\bf {\em K} fields, compactons, and thick branes }
\vglue 1  true cm

C. Adam$^{a*}$,   N. Grandi$^{\, b**}$, J. S\'anchez-Guill\'en$^{a\dagger}$,
and A. Wereszczy\'nski$^{c\dagger\dagger}$
\vspace{1 cm}

\small{ $^{a)}$Departamento de Fisica de Particulas, Universidad
     de Santiago}
     \\
     \small{ and Instituto Galego de Fisica de Altas Enerxias (IGFAE)}
     \\ \small{E-15782 Santiago de Compostela, Spain}
     \\ \small{ $^b)$ IFLP-CONICET }
     \\ \small{cc67 CP1900, La Plata, Argentina}
      \\ \small{ $^{c)}$Institute of Physics,  Jagiellonian
     University,}
     \\ \small{ Reymonta 4, 30-059 Krak\'{o}w, Poland}

\medskip
\end{center}

\normalsize
\vskip 0.2cm

\begin{abstract}
{\em K} fields, that is, fields with a non-standard kinetic term, allow for
soliton solutions with compact support, i.e., compactons. Compactons
in 1+1 dimensions may give rise to topological defects of the domain
wall type and with finite thickness in higher dimensions.
Here we demonstrate that, for an appropriately chosen kinetic term,
propagation of linear perturbations is completely suppressed outside the
topological defect, confining the propagation of particles inside the
domain wall. On the other hand, inside the topological defect the
propagation of linear perturbations is of the standard type, in spite of
the non-standard kinetic term. Consequently, this compacton domain wall
may act like a brane of finite thickness which is embedded in a higher
dimensional space, but to which matter fields are constrained.
In addition, we find strong indications that, when gravity is taken into
account, location of gravity in the sense of Randall--Sundrum works for
these compacton domain walls. When seen from the bulk, these finite thickness
branes, in fact, cannot be distinguished from infinitely thin branes.
\end{abstract}

\vfill

{\footnotesize
$^*$adam@fpaxp1.usc.es

$^{**}$grandi@fisica.unlp.edu.ar

$^{\dagger}$joaquin@fpaxp1.usc.es

$^{\dagger\dagger}$wereszczynski@th.if.uj.edu.pl }

\end{titlepage}

\section{Introduction}

In the last two decades the notion of a universe with more than
three space dimensions has excited some interest both in cosmology
and in theoretical high energy physics. One way to reconcile this
idea with the observational fact that only three space dimensions
are perceived is the assumption that the additional dimensions are
too small to be detected. Another possibility is to assume that
matter and (at least, non-gravitational) interactions are
restricted to a lower dimensional (concretely, three dimensional)
subspace. This subspace may be either strictly lower dimensional,
in which case the name ``three-brane'' has become customary. Some
recent reviews about these cosmological three-branes can be found,
e.g., in \cite{Lan1} - \cite{BrBr1}, which also serve as sources
for further references on the subject (which are too numerous to
be quoted here). Or the subspace may have a finite, although
probably very small, extension in the additional dimensions, that
is, it is of the topological defect type \cite{Akama1},
\cite{RuSh1}. In the latter case, the name ``thick brane'' has
been coined recently for this subspace on which the matter
propagation takes place (see, e.g., \cite{CEHS1} - \cite{AnVe1}
for some recent work). At least for a ``thick brane'', a dynamical
mechanism certainly has to be identified which provides the
confinement of all matter fields to the subspace. It is the
purpose of this letter to provide an alternative mechanism. Our
proposal bears some similarity to the work of others (e.g.
\cite{RuSh1}), in the sense that the ``thick brane'' is formed by
a topological defect, but the mechanism of confinement of the
matter fields is different. The basic idea is, in fact, very easy
to understand. Usually, the restriction of the propagation to a
topological defect is achieved by a potential which becomes very
strong outside this topological defect. But the propagation of a
field is generally determined by a balance between potential and
kinetic terms. Another possibility for restricting the propagation
to the topological defect is, therefore, a kinetic term which
becomes very small away from the defect, thereby enforcing that
the field takes its vacuum value in that region. This is precisely
what may happen in the case of {\em K} fields, that is, field
theories with a non-standard kinetic term.
\\
{\em K} fields already play a rather prominent role in cosmology,
where they offer a possible mechanism for early time inflation
(under the name of {\em K} inflation) \cite{APDM1} - \cite{COY1},
as well as a possible explanation both for the value of the
cosmological constant (problem of smallness, coincidence), and for
the late-time acceleration (\cite{APMS1} - \cite{babichev}), which
has been found to be a property of our universe in the last
decade. In the latter case, these models are nowadays known as
{\em K} essence theories.

\section{ The model}

To illustrate our proposal, we shall consider the simplest possible setting
where space time is 4+1 dimensional Minkowski space, and the topological
defect will be a simple domain wall which is effectively three dimensional.
The field theory is
given by the Lagrangian density
\be \label{lag}
{\cal L}= 4 | X | X - V(\xi)
\ee
where
\be
X \equiv \frac{1}{2}\partial_M \xi \partial^M \xi ,
\ee
\be
V(\xi ) \equiv
3 \lambda^4 (\xi^2 - a^2)^2 ,
\ee
$\xi$ is a scalar field, $\lambda$ and $a$ are positive constants,
and $M= 0 ,\ldots ,4$. Further, we use the Minkowski metric $g_{MN} =
{\rm diag} (1,-1,-1,-1,-1)$. The choice of the non-standard kinetic term
equal to
$4|X| X$ instead of $4X^2$ is not important for the purpose of this letter
(i.e., for the existence of the compacton solution and for the linear
perturbation analysis), but it {\em is} important for the global stability
of the field theory (\ref{lag}). For the kinetic term $4 X^2$, the energy is
not bound from below, see \cite{comp} for a detailed discussion.
\\
A first fact about the theory (\ref{lag}) which we need is that
when restricted to 1+1 dimensions, it has soliton solutions with
compact support. Indeed, choosing e.g. $x^4 \equiv y$, the theory
has the solution
\begin{equation}
\xi (y) = \left\{
\begin{array}{lc}
- a & \quad y \leq - \frac{\pi}{2\lambda }  \\
a \sin \lambda y & \quad
-\frac{\pi}{2\lambda } \leq y \leq
\frac{\pi}{2\lambda }  \\
a & \quad y \geq \frac{\pi}{2 \lambda },
\end{array}
\right. \label{compacton sol}
\end{equation}
which interpolates between the two distinct vacuum values $-a$ and
$a$, see \cite{comp} for details (for a more general discussion of
compactons, we refer to \cite{Arodz1} - \cite{Arodz3}). This
compacton solution is continuous and has continuous first
derivative. It is a domain wall solution in the 4+1 dimensional
Minkowski space.
\\
Remark: The compacton configuration has non-continuous second
derivative at the compacton boundary. In the field equations,
however, this discontinuity is multiplied by zero whenever it
shows up (i.e., multiplied by some power of the first derivative),
so that the expression for the field equation is continuous
everywhere.
\\
Remark: The Cauchy problem at the compacton boundary (i.e., for
the initial conditions $\xi (y_0)=a$, $\xi_y (y_0)=0$) is not
well-defined. The determining equation for $\xi_{yy}(y_0)$ is, in
fact, a cubic equation with the three roots $\xi_{yy}(y_0)=0,\pm
a\lambda^2$, corresponding to the vacuum, compacton and
anticompacton, respectively. Once this ambiguity is resolved, the
solution is unique in a finite neighborhood of $y_0$ (e.g., up to
the other boundary of the compacton). Observe that $y_0$ is
arbitrary due to translation invariance.
\\
In a next step we want to study the behavior of linear
perturbations about the domain wall (i.e., the compacton). Here
one simply inserts the field \be \xi (x^M) = \xi_0 (y) + \eta
(x^M) \ee (where $\xi_0$ is the compacton and $\eta $ is the
fluctuation) into the action of the theory (\ref{lag}). The
resulting linear equation for the fluctuation field for a general
Lagrangian density is (see \cite{comp}, \cite{bazeia}) \be
\partial_M ( {\cal L}_X \delta^M_N + {\cal L}_{XX} \xi^M_0 \xi^N_0 )\eta_N
- {\cal L}_{\xi\xi}\eta =0
\ee
where ${\cal L}_X \equiv \partial_X{\cal L}$, $\xi^M_0 \equiv g^{MN}
\partial_N \xi_0 $, etc.
Further, we already took into account that there are no
mixed terms in the Lagrangian, that is, ${\cal L}_{X\xi} =0$. In the
above expression the derivatives of the Lagrangian have to be evaluated for the
compacton field. For our model we have concretely
\be
{\cal L}_X = 8 |X| \, , \quad {\cal L}_{XX} = 8 \, {\rm sign} \, (X) \, , \quad
{\cal L}_{\xi\xi} = -12 \lambda^4 (3\xi^2 -a^2).
\ee
Taking into account that for the compacton $\xi_M =\delta_M^4 \xi_{x^4}$,
etc., we find the equation (where again $x^4 \equiv y$)
\be
-24 \, {\rm sign} \, (X) X_y \eta_y -24 |X| \eta_{yy} + 8 |X|
\partial^\mu \partial_\mu \eta
+12 \lambda^4 (3\xi_0^2 - a^2)\eta =0
\ee
where $X$, etc. have to be evaluated for the compacton field, and
$\mu = 0, \ldots ,3$.
In the region
outside the compacton, where $\xi_0$ takes its vacuum values $\pm a$, all
terms involving derivatives of the linear perturbation $\eta$ are multiplied
by zero, because $X=0$ and $X_y =0$ in that region. There we are left
with
\be
12 \lambda^4 (3\xi_0^2 - a^2)\eta =24 \lambda^4 a^2 \eta =0
\ee
which has the only solution $\eta =0$. There are, therefore, no linear
perturbations in that region, that is, all particle propagation is completely
suppressed.
\\
Inside the compacton we need the expressions
\be
\partial_y \xi_0 = a\lambda \cos \lambda y \, ,\quad
X= -\frac{1}{2} a^2 \lambda^2 \cos^2 \lambda y
\ee
and
\be
X_y = a^2 \lambda^3 \sin \lambda y \cos \lambda y
\ee
to arrive at the equation (after a division by $12a^2$)
\be
-\cos^2 \lambda y \,  \eta_{yy} +2 \lambda \sin \lambda y \cos \lambda y
\, \eta_y + \lambda^2 (3\sin^2 \lambda y - 1)\, \eta + \frac{1}{3}
\cos^2 \lambda y  \, \partial^\mu \partial_\mu \eta =0
\ee
or, after the variable change $z=\lambda y$,
\be
H \eta = -\frac{1}{3\lambda^2} \cos^2 z \, \partial^\mu
\partial_\mu \eta
\ee
where the differential operator $H$ is
\be \label{H-op}
H\equiv - \cos^2 z \, \partial_z^2 +2 \sin z \cos z \, \partial_z +3\sin^2 z
-1.
\ee
For a further evaluation, we use the separation of variables ansatz
$\eta =\bar \eta (z) \Phi (x^\mu)$, which leads to the equations
\be \label{H-eq}
H\bar \eta =\omega^2 \cos^2 z \, \bar \eta
\ee
and
\be \label{Phi-eq}
\partial^\mu \partial_\mu \Phi + 3\omega^2 \lambda^2 \Phi =0.
\ee
Before further discussing these equations, we have to determine the space of
functions on which the operator $H$ is supposed to act. We want the
perturbation $\eta$ to be continuous at the boundary of the compacton, i.e.,
at $z=\pm \pi /2$, therefore the space of functions is
 \begin{equation}
\bar \eta (z) = \left\{
\begin{array}{lc}
0 & \quad z \leq - \frac{\pi}{2}  \\
\sum_{n=1}^\infty b_n \cos nz & \quad
-\frac{\pi}{2} \leq z \leq
\frac{\pi}{2}  \\
0 & \quad z \geq \frac{\pi}{2} .
\end{array}
\right. \label{eta-z}
\end{equation}
Observe that the operator $ H$ maps the space of functions (\ref{eta-z})
into itself, so its action is well-defined on this space.
Observe also that the first derivative is not continuous at the
compacton boundary,
which is consistent with the fact that the compacton itself is continuous
together with its first derivative. Observe, finally, that there is no
discontinuous term in Eq. (\ref{H-eq}), because whenever a discontinuous
factor appears in that equation, it is multiplied by zero.
\\
It is easy to prove that the operator $H$ is positive
semi-definite on the space of functions (\ref{eta-z}), see Section
4.4 of Ref. \cite{comp}. It has, in fact, one zero mode $\bar
\eta_0 (z) =\cos z$, i.e., $ H\cos z =0$, and is positive definite
on the subspace $\eta = \sum_{n=2}^\infty b_n \cos nz$. Therefore,
Eq. (\ref{H-eq}) has the general solutions \be \label{H-sol} H\bar
\eta_n =\omega^2_n \cos^2 z \, \bar \eta_n \, , \quad \omega^2_0
=0 \, ,\quad \omega_n^2 >0 \, \, {\rm for} \, \, n\ge 1. \ee As a
consequence, the field equations (\ref{Phi-eq}) for the fields
$\Phi (x^\mu)$ on the domain wall (the ``thick brane'') are just a
collection of ordinary Klein--Gordon equations. There exists one
massless field due to the zero mode $\omega_0 =0$ in Eq.
(\ref{H-sol}), which is just the Goldstone field for the
spontaneously broken translational invariance in the $x^4$
direction. The other Klein--Gordon equations are for positive
square masses $m_n^2 = 3 \omega_n^2 \lambda^2 $, so there are no
tachyons on the brane. The propagation of perturbations on the
brane is, therefore, completely standard in spite of the
non-standard kinetic term of the model. The only way in which the
original, nonstandard theory enters at this stage is in the
determination of the values for the masses $m_n^2$, which depend
on the parameters of the original theory. The whole setting is, in
fact, quite similar to the reduction in the familiar Kaluza--Klein
case, in spite of the noncompact fifth dimension in our case.
Finally, the modes for nonzero $m_n^2$ can always be removed from
the physically accessible spectrum by choosing $\lambda$
sufficiently large.

\section{ Backreaction and localization of gravity}

In the sequel, we shall investigate the {\em K}-field equations in the
presence of gravity, and the possibility of localizing gravity in
compacton solutions like the one described above.
We will couple the scalar field to gravity minimally, and include the
dynamics of the gravitational sector in the form of a canonical 5D
Einstein term. The action is
\begin{equation}
  S=\int d^5x \sqrt {-g}\left( \kappa^{-2}(R - \Lambda)+
4 |X|X - V(\xi)\right)
\end{equation}
where $\Lambda$ is a cosmological constant and $X$ now includes the metric
\begin{equation}
  X = \frac 12 g^{MN}\partial_M \xi\partial_N\xi
\end{equation}
The resulting Einstein equations read
\begin{equation}
  \kappa^{-2}G_{MN}= 4|X|\partial_M\xi\partial_N\xi-\frac
12 g_{MN}\left(-\kappa^{-2}\Lambda+ 4 |X|X - V(\xi)\right)
\end{equation}
We will choose a 5D metric ansatz with a Minkowskian 4D slice, written in
the form
\begin{equation}
  ds^2=e^{-A(y)}\left(dt^2-d\vec x^{\,2}\right)-dy^2
\end{equation}
while for the scalar field we assume $\xi=\xi(y)$ as before.
The independent components of the Einstein equations now read
\begin{eqnarray}
&&\frac34 A_{yy} -  {A_y^{\, 2}} =
\frac13\left(\Lambda+\kappa^2 V(\xi)\right)
\nonumber \\
&&\frac34A_{yy} = \kappa^2\xi_y^{\, 4} .
\label{system}
\end{eqnarray}
The next step would be to solve the boundary value problem for the
above coupled system (\ref{system}) to prove that the compacton
solution is not spoiled by the gravitational backreaction. Since
the solution of the full system will be more complicated, we will
just assume in a first step that the solution exists and explore
the effects of the compacton on the gravitational degrees of
freedom.
\\
By definition, outside the support of the compacton the {\em K} field $\xi$ is
in its vacuum $\xi=\pm a$. There the equations for $A$ take the form
\begin{eqnarray}
A_y^{\, 2}&=& -\frac13 \Lambda
\nonumber \\
A_{yy}&=&0
\end{eqnarray}
that immediately imply the vacuum $AdS$ solution $A=
\sqrt{- \Lambda /3}\,|y|+constant\equiv \bar A$. This represents exactly the
same bulk solution as in the case on an infinitely thin brane \cite{RaSu1}.
\\
The metric perturbation analysis for the case of a thick brane has
been performed in \cite{CEHS1}. The main result is that the 4D
graviton decouple from scalar field perturbations and that, to see
wether there is localization of gravity, one should prove that the
graviton wave function $e^{-A}$ is normalizable. In other words we
should verify that
\begin{equation}
  \int dy \, e^{-A}<\infty
\end{equation}
This is obviously satisfied for the above solution, provided it is completed
inside the compacton with an integrable function.
The corresponding 4D Plank mass reads
\begin{equation}
M_4^2 =\kappa^{-2} \int dy \, e^{-A}
\end{equation}
Separating the above integral into its contributions inside the compacton
and outside it, using the fact that the exterior solution for $A$ is that
of an infinitely thin brane $\bar A$, and adding and subtracting the interior
contributions for $\bar A$, we get
\begin{equation}
M_4^2 = \bar M_4^2+\kappa^{-2} \int_{int} dy \left(e^{-A}-\,e^{-\bar A}\right)
\end{equation}
where $\bar M_4$ is the 4D Plank mass for an infinitely thin brane with the
same bulk solution. The function in parenthesis vanishes at the boundary of
the compacton, and its first derivative also vanishes. Its second derivative
is negative for positive $y$ and positive for negative $y$. The only possible
conclusion is that the function itself is negative inside the support, giving
a negative contribution to the effective four dimensional Plank mass.
\\
The conclusion is that, if the compacton solution still exists in the
backreacting system, it represents a finite thickness brane-world that is
indistinguishable from an infinitely thin brane when it is seen from the
bulk. The observer on the brane, on the other hand, measures localized 4D
gravity with a Plank mass suppressed with respect to that of the infinitely
thin case.
\\
We still should comment on the solution of the full system and on
the related stability of the compacton under gravitational
backreaction. First of all, it still remains true that for vacuum
boundary conditions (i.e., for $\xi (y_0) =a$, $\xi_y (y_0)=0$),
the second derivative is not uniquely defined but, instead, obeys
a cubic equation with the three roots $\xi_{yy}(y_0) =0,\pm
a\lambda^2$. This indicates that it should still be possible to
join the vacuum with the compacton or anticompacton boundary,
respectively. Secondly, both a power series expansion about the
compacton boundary and a numerical integration from a point very
near the compacton boundary (i.e., with the inclusion of the
second derivative $\xi_{yy}(y_0)$ in order to have a well-defined
solution) up to the center of the compacton (a point $y_1$ such
that $\xi (y_1)=0$) lead to very reasonable results. The radius of
the compacton $\vert y_1 - y_0 \vert$ is very similar to the case
without gravitation for sufficiently small values of the
cosmological constant $\Lambda$ and gravitational coupling
$\kappa$. Further, $\Lambda$ tends to increase the compacton
radius, whereas $\kappa$ tend to shrink it, in complete agreement
with general expectations. A detailed analytical and numerical
discussion of the full K field and gravitation system is beyond
the scope of this letter and shall be presented elsewhere.

\section{ Discussion}

We have proposed a simple and efficient mechanism for the
production of thick branes, that is, topological defects within a
higher dimensional space, to which the propagation of linear
perturbations is confined. These thick branes have the interesting
property that they are of strictly finite extension in the
additional dimension. The main ingredients of the proposal are the
use of a model with a non-standard kinetic term and the
observation that topological defects with a compact support
(compactons) exist in such models. The total suppression of the
propagation of fields outside the support of the compacton is an
automatic result of the model. Furthermore, propagation inside the
topological defect (i.e., inside the brane) is standard in spite
of the non-standard kinetic term. Specifically, there are no
tachyons on the brane, and the evolution of linear perturbations
is both unitary and causal. Inside the brane, the only remaining
effect of the original K field theory resides in the values of the
masses of the (Klein--Gordon type) linear fluctuation field. A
study of the dynamical evolution of the full system, i.e., the
inclusion of time dependence would therefore be very interesting,
in order to discriminate the resulting physics on the brane from
other scenarios.
\\
Let us emphasize that in this letter our main purpose was to
present the generic mechanism of thick brane generation via {\em
K} fields and compactons. For a possible use of this idea in
cosmological or particle physics considerations, additional
structures have to be added. First of all, the existence of
compacton solutions and the suppression of propagation in a vacuum
background is a rather generic feature of {\em K} field theories.
All that is needed is that the kinetic term remains non-standard
in a specific way for fields near their vacuum value \cite{comp}.
There exists, therefore, a large class of {\em K} field theories
which show essentially the same features. Secondly, it will be of
interest to add fermions to the theory, which probably give rise
to the presence of fermionic zero modes, as in the case of a
standard background topological defect \cite{RuSh1}. This may also
open the way for introducing supersymmetry for K field theories.
Thirdly, the system with gravity included should be further
analysed. We already found that, provided the compacton domain
wall is not destabilized by the addition of gravity, bulk gravity
solutions of the Randall--Sundrum type (that is, localization of
gravity on the brane) do exist. Further, we found strong
indications for the stability of the full system with gravity
included. Still, this latter issue should be investigated in more
detail. A fourth issue not touched in this letter is the question
whether theories with a non-standard kinetic term of the type
required for the existence of compactons may be induced as
effective low energy theories from some more fundamental theories
at higher energies. These and many more problems are subject to
further investigations.

\section*{Acknowledgements}

C.A. and J.S.-G. thank MCyT (Spain) and FEDER
(FPA2005-01963), and support from Xunta de Galicia (grant
PGIDIT06PXIB296182PR and Conselleria de Educacion). A.W. is
partially supported from Jagiellonian University (grant WRBW
41/07).
N.E.G. thanks  Xunta de Galicia for support and Departamento
de Fisica de
Particulas - Universidade de Santiago de Compostela for hospitality
during this work.

\end{document}